\documentclass[aps,twocolumn,raggedbottom,prl,amsmath,showpacs,nobalancelastpage,amssymb,groupedaddress]{revtex4}
\usepackage{epsfig}
\begin{document}
\title{Shape resonances for ultracold atom gases in 
carbon nanotube waveguides}
\author{V. Peano} \author{M. Thorwart} \author{C. Mora} \author{R. Egger}
\affiliation{Institut f\"ur Theoretische Physik, 
Heinrich-Heine-Universit\"at, D-40225 D\"usseldorf, Germany}

\date{\today}

\begin{abstract}
We propose an experimentally viable setup for the realization
of  one-dimensional ultracold atom gases in a
nanoscale magnetic waveguide formed by two 
doubly-clamped suspended carbon nano\-tubes.  
All common decoherence and atom loss mechanisms are shown to 
be small. We discuss general consequences of a non-parabolic
confinement potential, in particular novel two-body shape resonances,
which could be observed in this trap. 
\end{abstract}
\pacs{03.65.Nk, 03.75.Gg, 73.63.Fg}
\maketitle

The ongoing progress in the fabrication and manipulation of 
micro- or nanoscale structures has recently allowed
for systematic studies of ultracold atom gases,
where current-carrying wires generate magnetic fields trapping 
neutral atoms (`atom chips')  \cite{Folman02,Reichel02}. For instance,
the Bose-Einstein condensation (BEC) of microchip-confined atoms 
has been successfully demonstrated by several groups \cite{atomBEC}.
Such an approach is particularly promising in the context
of integrated atomic matter-wave interferometry and optics \cite{Kasevich}, 
and combines the strengths of nanotechnology and atomic physics.
So far, deco\-her\-ence and atom loss constitute  central im\-pe\-di\-ments, 
since atoms are relatively close to `hot' macroscopic surfaces
or current-carrying wires (with typical diameters of several $\mu$m),
where the Casimir-Polder potential and Johnson noise can seriously affect 
stability  \cite{Henkel99,chin,Schroll03}. 
To reduce these effects, further miniaturization would be desirable.
While at first sight this goal conflicts with the requirement of large 
currents forming the trapping potentials, we propose that when
using suspended carbon nanotubes (NTs) \cite{tubes} (with 
diameters of a few nm) as  wires,
nanoscale atom chip devices basically free of decoherence
or atom loss can be built with state-of-the-art technology. 
With relevant length scales below optical and cold-atom de
Broglie wavelengths, this also paves the way for the observation
of interesting and largely unexplored many-body physics 
in one dimension (1D) \cite{Petrov04}, involving either bosons or fermions.  
Examples include the interference properties of interacting matter waves 
\cite{chen}, spin-charge separation \cite{recati}, fractional statistics
and atom number fractionalization \cite{pham}, and the 1D analogue of the 
BEC-BCS crossover  \cite{becbcs}.
Previous realizations of 1D cold atoms
were reported using optical lattices \cite{esslinger,Paredes04,Weiss} 
and magnetic traps \cite{Goerlitz01}, but they
involve arrays of 1D or elongated 3D systems,
where the above many-body effects are difficult to observe.
Our proposal completely eliminates unwanted substrate effects 
and implies a drastically reduced transverse size (a few nm)
of the cloud. Hence rather high atom densities could
be achieved, allowing for stable operation and 
sensible optical detection schemes.

\begin{figure}[t]	
\begin{center}
\epsfig{figure=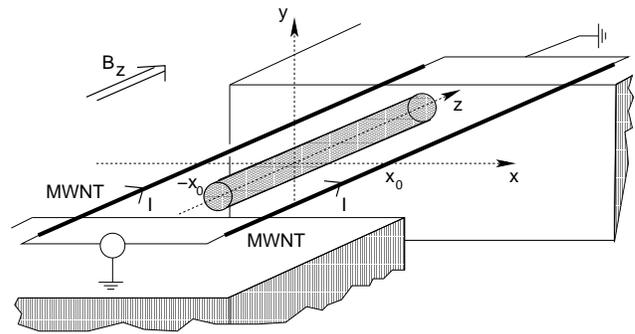,width=83mm,keepaspectratio=true}
\caption{Sketch of the proposed device. The two current-carrying
suspended NTs are positioned at $(\pm x_0,0,z)$. 
The shaded region indicates the atom gas. \label{fig.1}}
\end{center}
\end{figure}

The proposed nanoscale waveguide confining ultracold atoms to 
1D is sketched in Fig.~\ref{fig.1}. 
The setup employs two suspended doubly-clamped NTs, where 
nanofabrication techniques routinely allow for trenches with 
typical depth and length
of several $\mu$m \cite{tubes}. To minimize decoherence and loss effects
\cite{chin},
the substrate should be insulating apart from thin metal strips 
to electrically contact the NTs.
 Since strong currents in the mA-regime  are necessary, thick
 multiwall nanotubes (MWNTs) or `ropes' (bundles) \cite{tubes} are best suited. 
Due to the suspended geometry, the disturbing influence of the 
substrate is largely eliminated, and 
employing an additional longitudinal magnetic field
to eliminate Majorana spin flips \cite{sukumar,jones},
neutral atoms in a weak-field seeking state can be 
trapped between the NTs.
Studying various sources for decoherence, heating or atom loss,
and estimating the related time scales, we find
that, for reasonable parameters, detrimental effects are small.
We also analyze effects due to the non-parabolicity of the
transverse trap potential, which is inevitable in all common traps
but particularly pronounced in the present case.  We find
many two-body shape resonances, which allow to tune atom-atom 
interactions similar to standard Feshbach resonances.  
Previous work has only studied parabolic traps,
where center-of-mass (COM) variables decouple. Then
incoming scattering waves can visit only one out of many bound states 
normally available, leading to  a single `confinement-induced resonance' 
\cite{olshanii}.  
Here we solve the two-body problem for general transverse 
confinement, and then apply the results to the proposed trap.
As a concrete example, we shall consider $^{87}$Rb atoms in the 
weak-field seeking hyperfine state $|F,m_F\rangle=|2,2\rangle$.

\begin{figure}[t]
\begin{center}
\epsfig{figure=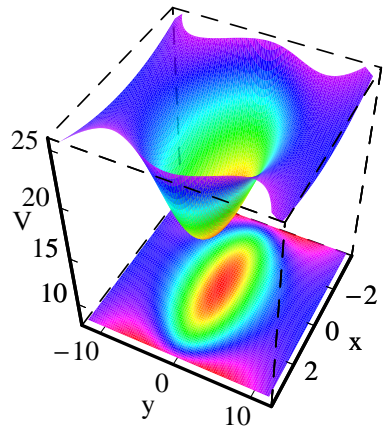,width=50mm,keepaspectratio=true,angle=0}
\caption{(Color online) Transverse trapping potential of the nanoscale 
waveguide, see text. \label{fig.2}}
\end{center}
\end{figure}

We next describe the setup in Fig.~\ref{fig.1} in detail, where
the same (homogeneous) current $I$ flows through the
two NTs at $(\pm x_0,0,z)$, and a magnetic field $B_z$
is applied along the $z$-direction.  Neglecting boundary
effects due to the finite tube length $L$,
the magnetic field at position ${\bf x}=(x,y,z)=({\bf x}_\perp,z)$ is 
\begin{eqnarray*}
\mathbf{B}({\bf x})&=&\frac{\mu_0 I}{2\pi}
\frac{1}{[(x-x_0)^2+y^2][(x+x_0)^2+y^2]}
  \\&&
\times\left( \begin{array}{c} -2y(x^2+x_0^2+y^2)\\ 2x(x^2+y^2-x_0^2)\\ 0\\
\end{array}\right) +\left( \begin{array}{c} 0\\0\\B_z\\\end{array}\right)
\end{eqnarray*}
with the vacuum permeability $\mu_0$.   
The transverse confinement potential is
$V({\bf x}_\perp)=\mu |{\bf B({\bf x})}|$, where $\mu=m_F g_F \mu_B$ 
 with the Land{\'e} factor $g_F$ and the Bohr magneton $\mu_B$.  
Under the adiabatic approximation \cite{sukumar},
$m_F$ is a constant of motion, and the
potential is harmonic very close to the minimum of the trap, with
frequency $\omega=  [\mu/(m B_z)]^{1/2} \mu_0 I / (\pi x_0^2)$ 
and associated transverse confinement length $a_\perp=(\hbar/m\omega)^{1/2}
\ll x_0$, where $m$ is the atom mass.  
The adiabatic approximation is valid as long as $\omega\ll \omega_L$
 with the Larmor frequency $\omega_L= \mu B_z/\hbar$. 
Trapped atoms can in general also make non-adiabatic
Majorana spin flips to a strong-field seeking state,
and thereby escape from the trap \cite{Folman02,jones}. 
The associated loss rate is $\Gamma\simeq (\pi\omega/2)\exp(1-1/\chi)$
with $\chi=\hbar\omega/(\mu B_z)$ \cite{sukumar}. 
For convenience, we switch to a dimensionless form of the full potential
$V({\bf x}_\perp)$ by measuring energies in units of $\hbar \omega$ and 
lengths in units of $a_\perp$, 
\begin{equation} \label{fullpot}
\chi V= \left( 1+\chi d^4 \frac { (x^2+y^2)
 [ (x^2+y^2+d^2)^2-4x^2d^2] }
{ [(x-d)^2+y^2]^2[(x+d)^2+y^2]^2 } \right)^{1/2},
\end{equation}
which depends only on $d= x_0/a_\perp$ and $\chi$.
To give an example, we take $I=1$~mA, representing a 
reasonable current through thick MWNTs \cite{tubes}.  For $d=10$ and 
$\Gamma/\omega = 10^{-6}$  ($\chi=0.067$), we obtain $B_z= 20$~G,
$x_0=79$~nm and  $\omega=2\pi \times 1.85$~MHz, 
corresponding to a very tight trap.
The potential is shown for these parameters in Fig.~\ref{fig.2} \cite{nota2}.  

For stable operation, it is essential that destructive effects 
like atom loss, heating or decoherence are small. 
One loss process proceeds via (i) non-adiabatic Majorana spin flips as
discussed above.
Another one is (ii) atom loss due to tunneling out of the trap.
The WKB tunneling rate $\gamma_{\rm t}$
can be estimated easily for an atom in the transverse ground state
escaping along the least-confined $y$-direction, see Fig.~\ref{fig.2},
\[
\gamma_{\rm t}/\omega \simeq (2\pi)^{-1}\exp\left[ -2 \sqrt{2} \int_{y_1}^{y_2} 
dy [V(0,y)-1]^{1/2}\right].
\]
For the above parameters, we find numerically $y_1=1.47, y_2=68.09$,
and hence $\gamma_{\rm t}/\omega\approx 10^{-100}$.
Atom loss may also originate from (iii) noise-induced spin flips,
where current fluctuations cause a fluctuating magnetic field 
generating the Majorana spin flip rate \cite{Henkel99} 
\[
\gamma_{\rm sf} \simeq  \left(\frac{\mu_0\mu}{2\pi\hbar
 x_0}\right)^2 \frac{S_I(\omega_L)}{2}, \
S_I(\omega)=\int dt e^{-i\omega t}\langle I(t)I(0)\rangle.
\]
At room temperature and for typical voltages $V_0\approx 1$~V, we have
$\hbar\omega_L\ll k_BT\ll eV_0$,  and $S_I(\omega_L)$ is
expected to equal the shot noise $2eI/3$ of a diffusive wire. 
For the parameters above,  a rather small escape rate results,
$\gamma_{\rm sf}\approx 0.017$~Hz. 
Next we study (iv) the transverse  NT deflection 
 due to their mutual magnetic repulsion, using 
a standard elasticity model for a doubly clamped wire 
in the limit of small deflections \cite{Sapmaz03}. 
For small NT displacements $\phi(z)$, 
the equation of motion is $\rho_L \ddot{\phi}=-Y M_I \phi'''' + \mu_0
I^2/(4 \pi x_0)$, where $\rho_L$ is the linear mass density, 
$Y$ is Young's modulus and $M_I$ the NT's moment of inertia.
The static solution under the boundary conditions 
$\phi(0,L) = \phi'(0,L)=0$ 
is  $\phi(z) =\mu_0 [Iz(z-L)]^2/(96 \pi Y M_I  x_0)$. 
Using typical material parameters from Ref.~\cite{tubes},
the maximum displacement is $\phi(L/2)\approx 0.08$nm 
for $L=10\mu$m.  Hence the mutual magnetic repulsion of the NTs is weak. 
(v) Thermal NT vibrations might create decoherence  and heating, 
and could even cause a transition to the first excited state of the 
trap.   The maximum mean square displacement is $\sigma^2=\langle 
\phi^2(L/2)\rangle=k_B T L^3/(192 Y M_I)$ \cite{Sapmaz03}, which for the above
 parameters
gives $\sigma\approx 0.2$~nm at room temperature. This 
is much smaller than the transverse size $a_\perp$ of the atomic cloud.
Detailed analysis shows that the related decoherence rate is also negligible. 
Another decoherence mechanism comes from (vi) current fluctuations
 in the NTs. Following the analysis of Ref.~\cite{Schroll03}, the 
corresponding decoherence rate can be estimated for the above
parameters as $\gamma_{\rm c}/\omega < 10^{-7}$.  We conclude that no 
serious decoherence, heating or loss mechanisms are expected for 
reasonable parameters of this nanotrap.

Next we discuss $s$-wave atom-atom interactions in this trap.  We assume that 
the 3D interaction can be described by a Fermi pseudopotential \cite{huang}
via the 3D scattering length $a$.  
To keep generality, let us investigate two-body scattering for arbitrary 
confining potential $V({\bf x}_\perp)$, and later specialize to 
Eq.~(\ref{fullpot}).
Unlike for a harmonic trap, the problem does {\sl not}\ 
decouple when formulated in terms of the COM coordinate
${\bf R}=({\bf x}_1+{\bf x}_2)/2=({\bf R}_\perp,Z)$
and the relative coordinate ${\bf r}={\bf x}_2-{\bf x}_1 
= ({\bf r}_\perp, z).$ Following standard steps \cite{Petrov04}, 
the pseudopotential is enforced as a boundary condition for the 
two-particle wavefunction,
\begin{equation}\label{bc}
\Psi({\bf R},{\bf r}\to 0) = \frac{f({\bf R})}{4\pi |{\bf r}|}(1-|{\bf r}|/a).
\end{equation}
Without loss of generality, we set the longitudinal COM
momentum to zero such that $Z$ drops out. Furthermore, 
we can put ${\bf r}_\perp =0$ and then let $z\to 0$ in Eq.~(\ref{bc}).  
The solution of the two-particle Schr\"odinger equation 
$\Psi({\bf R}_\perp,z)$ then takes the general form
\begin{equation}\label{schr2}
 \Psi({\bf R}_\perp,z) = \Psi_0({\bf R}_\perp,z) 
 +\!\int\! d{\bf R}^\prime_\perp
 G_E({\bf R}_\perp,z;{\bf R}_\perp^\prime,0)\frac{f({\bf R}_\perp^\prime)}{m}.
\end{equation}
For $\Psi_0=0$, this leads to bound-state solutions with energy 
$E<2E_0$ discussed elsewhere \cite{later}, 
where $E_0$ is the single-particle ground-state energy.
Here we focus on scattering solutions at low energies $E$ slightly 
above $2E_0$, where exactly one transverse channel is open.  Then
 $\Psi_0({\bf R}_\perp,z)=e^{ikz}  \psi^2_0({\bf R}_\perp)$ describes
two incoming atoms with (small) relative longitudinal momentum 
$\hbar k=\sqrt{2m(E-2E_0)}$ in the (transverse) single-particle state
$\psi_0({\bf R}_\perp)$ with energy $E_0$. 
Some algebra yields the two-particle Green's function 
\begin{eqnarray}
&& G_E({\bf R}_\perp,z;{\bf R}^\prime_\perp,0)=\psi^2_0({\bf R}_\perp)
\bar{\psi}^2_0 ({\bf R}_\perp^\prime) \frac{im}{2k}e^{ik|z|} 
\nonumber\\
&+&\int_0^\infty dt\,e^{Et}\sqrt{\frac{m}{4\pi t}}e^{-z^2m/4t}
\tilde{G}_t({\bf R}_\perp,{\bf R}_\perp^\prime),\label{barnes} 
\end{eqnarray}
where the bar denotes complex conjugation, and
\begin{eqnarray*}
\tilde{G}_t({\bf R}_\perp,{\bf R}_\perp^\prime)
&=& [G_t^{(0)}]^2
-e^{-2E_0t}\psi^2_0({\bf R}_\perp)\bar{\psi}^2_0({\bf R}_\perp^\prime), \\
G^{(0)}_t({\bf R}_\perp,{\bf R}_\perp^\prime)&=&
\sum_{\lambda} e^{-E_\lambda t}\psi_\lambda({\bf R}_\perp)
\bar{\psi}_\lambda({\bf R}_\perp^\prime).
\end{eqnarray*}
Here, $G^{(0)}_t$ is the (transverse) single-particle Green's function,
with single-particle states $\psi_\lambda$ and energy $E_\lambda$,
and the two-particle Green's function $\tilde{G}_t$ 
acts on the  Hilbert space ${\cal H}_{\rm closed}$,
orthogonal to the open channel.
Inserting Eq.~(\ref{barnes}) into Eq.~(\ref{schr2})
yields for $|z|\rightarrow\infty$ a standard scattering solution
$\Psi({\bf R},z)=\psi^2_0({\bf R}_\perp)(e^{ikz}+f_e(k)e^{ik|z|})$
with scattering amplitude
\begin{equation}\label{sa}
f_e(k)=\frac{i}{2k}\int d{\bf R}^\prime_\perp \bar{\psi}^2_0({\bf R}_\perp^\prime)
f({\bf R}_\perp^\prime).
\end{equation}
Enforcing Eq.~(\ref{bc})  then leads to an integral equation,
\begin{eqnarray} \label{gastone} 
&& -\frac{f({\bf R}_\perp)}{4\pi a}= 
 \int d{\bf R}_\perp^\prime\zeta_E({\bf R}_\perp,{\bf R}_\perp^\prime)
f ({\bf R}_\perp^\prime)
\\ \nonumber
&&+\psi^2_0({\bf R}_\perp)
 +\frac{i\psi^2_0({\bf R}_\perp)}{2k} \int d{\bf R}_\perp^\prime
\bar{\psi}^2_0({\bf R}_\perp^\prime) f({\bf R}_\perp^\prime),
\end{eqnarray}
where the singular behavior has been split off by
introducing a regularized kernel in ${\cal H}_{\rm closed}$, 
\[
\zeta_E=\int_0^\infty\!\!\!\frac{dt}{\sqrt{4\pi m t}}
\left(e^{Et}\tilde{G}_t({\bf R}_\perp,{\bf R}_\perp^\prime)
-\frac{m}{4\pi t}\delta({\bf R}_\perp-{\bf R}_\perp^\prime) \right).
\]
The integral equation (\ref{gastone}) is most conveniently solved by 
expansion in a suitable orthonormal basis $|j\rangle$, 
\begin{equation}\label{deff}
|f\rangle = \sum_j f_j |j\rangle, \quad
 f_j = \int d{\bf R}_\perp \langle j|{\bf R}_\perp
\rangle f({\bf R}_\perp),
\end{equation}
where $|0\rangle$ corresponds to
$\langle {\bf R}_\perp|0\rangle= c\psi^2_0({\bf R}_\perp)$ 
with normalization constant $c$. 
Thereby we can express Eq.~(\ref{gastone}) in compact notation, 
\begin{equation} \label{ham}
-\frac{|f\rangle}{4\pi a}=\frac{|0\rangle}{ c}+\frac{i}{2k}
\frac{|0\rangle}{c^2}\langle0|f\rangle+\zeta_E |f\rangle,
\end{equation}
which is solved by 
\begin{equation}\label{solution}
|f\rangle= \frac{-1/c}{1-i/(k a_{\rm 1D})}
\left(\zeta_E+\frac{1}{4\pi a}\right)^{-1}|0\rangle.
\end{equation}
Here, the parameter $a_{\rm 1D}$ follows in the form
\begin{equation}\label{cl}
a_{\rm 1D}=-\frac{2c^2}{\langle 0| [\zeta_{E}+1/(4\pi a)]^{-1}|0\rangle}.
\end{equation}
Since $f_e(k)=-1/(1+ik a_{\rm 1D})$ follows from Eq.~(\ref{sa}),
$a_{\rm 1D}$ can be identified with the {\sl 1D scattering length}.
The effective 1D atom-atom interaction potential is then $V_{\rm 1D}(z,z')
=g_{\rm 1D} \delta(z-z')$ with interaction strength
 $g_{\rm 1D}=-2 \hbar^2/(m a_{\rm 1D})$ \cite{Petrov04,olshanii}. 
For very low energies, $k\to0$, we can now put $E=2E_0$ in 
Eq.~(\ref{cl}). For a binding trap, $\zeta_{2E_0}$ is an Hermitian operator 
with discrete spectrum $\{\lambda_n\}$ and eigenvectors $|e_n\rangle$,
and hence we find
\begin{equation}\label{u0}
a_{\rm 1D}^{-1} = -\frac{1}{2c^2}
 \sum_n \frac{|\langle 0|e_n \rangle|^2}{\lambda_n+1/(4\pi a)}.
\end{equation}
Let us then denote $H_{\rm closed}$ as the projection of 
the Hamiltonian to ${\cal H}_{\rm closed}$. The boundary condition for  the 
corresponding  
bound state is enforced by $-|f\rangle/(4\pi a)=\zeta_E |f\rangle$.
 Thus, for $a=-(4\pi \lambda_n)^{-1}$,  a bound state of 
$H_{\rm closed}$ with energy $E=2E_0$ exists, 
fulfilling Eq.~(\ref{bc}) with $f({\bf R})=\langle{\bf R}_\perp|e_n\rangle$. 
In agreement with general arguments \cite{olshanii,timmer},
provided  $\langle 0|e_n\rangle\neq 0$, 
a bound state of $H_{\rm closed}$ at $a=-(4\pi\lambda_n)^{-1}$ corresponds to 
 a {\sl shape resonance} in the 1D interaction strength
 $g_{\rm 1D}$.

\begin{figure}[t]
\begin{center}
\epsfig{figure=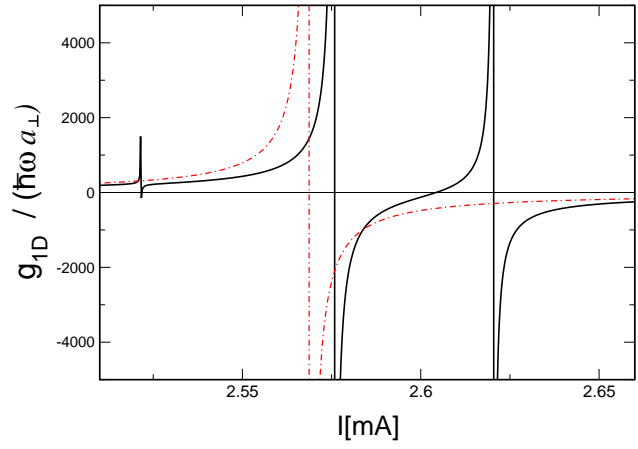,width=83mm,keepaspectratio=true,angle=0}
\caption{(Color online) The solid curve gives $g_{\rm 1D}$
 as function of $I$ for the potential 
(\ref{fullpot}) via numerical solution of Eq.~(\ref{u0}) for
$d\simeq 13$ and $\chi=0.067$.
The corresponding parabolic prediction for the same 
$\omega$ is shown as dashed curve.
\label{fig.3}}
\end{center}
\end{figure}

The two-body problem can thereby be solved 
numerically for any given confinement potential by diagonalizing 
$\zeta_{2E_0}$.  We  note in passing that for certain potentials, 
including a hard-box confinement,  
Eq.~(\ref{u0}) can also be computed analytically in closed form 
\cite{later}.   In general, every eigenvalue $\lambda_n$  corresponds to a 
different shape resonance, unless the overlap $\langle 0|e_n\rangle$
vanishes due to some underlying symmetry. For a parabolic confinement, 
the decoupling of the
COM motion implies that only one resonance is permitted, and 
some algebra gives indeed
 from Eq.~(\ref{u0}) the known result $a_{\rm 1D}=-(a^2_\perp/a) 
(1-1.0326 a/a_\perp)$ \cite{olshanii}.
For a general confinement, there are in principle infinitely many resonances. 
In practice, however, only few of them can be resolved since
 most of them appear at $a/a_\perp\to 0$. In addition, most resonances
 are extremely sharp and will therefore be hard to detect. 
The 1D interaction strength $g_{\rm 1D}$ 
for the potential (\ref{fullpot}) in the case of ${}^{87}$Rb atoms is shown in
Fig.~\ref{fig.3} as a function of the NT current $I$.
At least two different shape resonances are predicted to be observable
for realistic and discernible values of $I$ in the mA regime.
These resonances could be used to tune the strength and the sign
of the two-body interactions by simply adjusting the NT current. 
 
To conclude, we propose a nanoscale waveguide for ultracold atoms 
based on doubly clamped suspended nanotubes. 
All common sources of imperfection can be made sufficiently
small to enable stable operation of the setup.
Detection certainly constitutes an experimental challenge in this
truly 1D limit.  However, we note that 
single-atom detection schemes are currently being developed, 
which would also allow to probe the tight 1D cloud here,
e.g., by combining cavity quantum electrodynamics 
with chip technology \cite{Reichel02}, or by using additional perpendicular
wires/tubes `partitioning' the atom cloud \cite{reichel}.
This may then allow to study interesting many-body physics in 
1D in an unprecedented manner. 
Atom-atom interactions can be tuned by shape resonances,
 which we have described for arbitrary transverse confinement. 

We thank A. G\"orlitz and A.O. Gogolin for discussions,
and acknowledge support by the SFB/TR 12.

\begin{thebibliography}{99}
\bibitem{nota2} 
In order to create a 1D trap, a weak longitudinal confinement
should be added, e.g., by means of additional
NTs perpendicular to the $z$-axis (at $y=0$), see Ref.~\cite{reichel}.
\end{thebibliography}
\end{document}